\documentclass{article}
\usepackage{spconf,amsmath,amssymb,graphicx}

\usepackage[T1]{fontenc}
\usepackage[utf8]{inputenc}
\usepackage{multirow}
\usepackage{microtype}
\usepackage{booktabs}
\usepackage{xcolor}
\usepackage{hyperref}
\usepackage{comment}

\definecolor{improvement}{rgb}{0.0, 0.5, 0.0}

\newcommand{\inlinepar}[1]{\vspace{0.5\baselineskip}\noindent\textbf{#1}\hspace{0.5em}}

\title{FlowW2N: Whispered-to-Normal Speech Conversion via Flow-Matching}

\makeatletter
\gdef\@name{{\em Fabian Ritter-Gutierrez$^{1}$, Md Asif Jalal$^{1}$, Pablo Peso Parada$^{1}$, Karthikeyan Saravanan$^{1}$,}\\
            {\em Yusun Shul$^{2}$, Minseung Kim$^{2}$, Gun-Woo Lee$^{2}$, Han-Gil Moon$^{2}$}\\}
\makeatother
\address{$^{1}$Samsung Electronics R\&D Institute UK (SRUK), London, United Kingdom\\
         $^{2}$Samsung Electronics, Mobile eXperience Business, Suwon, Republic of Korea}

\begin{document}
\ninept
\maketitle

\begin{abstract}
Whispered-to-normal (W2N) speech conversion aims to reconstruct missing phonation from whispered input while preserving content and speaker identity. This task is challenging due to temporal misalignment between whisper and voiced recordings and lack of paired data. We propose FlowW2N, a conditional flow matching approach that trains exclusively on synthetic, time-aligned whisper-normal pairs and conditions on domain-invariant features. We exploit high-level ASR embeddings that exhibits strong invariance between synthetic and real whispered speech, enabling generalization to real whispers despite never observing it during training. We verify this invariance across ASR layers and propose a selection criterion optimizing content informativeness and cross-domain invariance. Our method achieves SOTA intelligibility on the CHAINS and wTIMIT datasets, reducing Word Error Rate by 26-46\% relative to prior work while using only 10 steps at inference and requiring no real paired data, validated by a subjective listening study and F0-contour analysis.
\end{abstract}


\section{Introduction}
\label{sec:intro}

Whispered speech is characterized by the absence of vocal fold vibration \cite{whatwhisper}, resulting in an acoustic signal devoid of a fundamental frequency ($F_0$) and the corresponding harmonic structure~\cite{acoustics_of_whisper}, which impedes intelligibility and naturalness. Whispered-to-Normal (W2N) speech conversion computationally reconstructs these missing acoustic features, transforming whispered input into voiced speech while preserving linguistic content and speaker identity.

Unlike speech denoising~\cite{sgmse,2023storm}, where source and target remain aligned at the content level, W2N involves phonetic and speaking-rate mismatches~\cite{dataaug-whisper-asr}: speakers alter phonetic durations when switching between whispered and voiced phonation~\cite{Morris2002ReconstructionOS, acoustics_of_whisper}, so paired recordings are temporally misaligned.

Prior W2N methods include VAE-based approaches~\cite{kameoka2019acvae} that suffer from over-smoothing and GAN-based methods~\cite{wagner2024gancomparative, vocoderfreevcwhisper, nonparallelvcw2n} prone to training instability and audible artifacts. Recent SSL-based approaches such as WESPER~\cite{rekimoto2023wesper} and DistillW2N~\cite{tan2025distillw2n} leverage HuBERT Soft~\cite{hsu2021hubert,van2022comparison} features. A persistent limitation across these methods is degraded intelligibility: converted-speech WER remains substantially higher than that of the input whisper.

Diffusion~\cite{ho2020denoising, kong2020diffwave} and flow-matching~\cite{lipman2022flowmatching} models have advanced high-fidelity speech synthesis \cite{guo2024voiceflow,ren2025reflowvc}. Flow matching~\cite{lipman2022flowmatching,liu2022rectified} learns a velocity field transporting samples from a source to a target distribution via linear interpolation among paired samples. However, applying it directly to W2N fails: the temporal misalignment above causes the interpolated trajectory to blend distinct phonemes from disparate time steps -- a ``phoneme boundary blur'' effect that prevents learning a meaningful vector field, and which we show in Section~\ref{sec:paired_limitation} that standard alignment techniques such as DTW~\cite{dtw} do not resolve.

\begin{figure*}[t]
    \centering
    \includegraphics[width=0.6\textwidth]{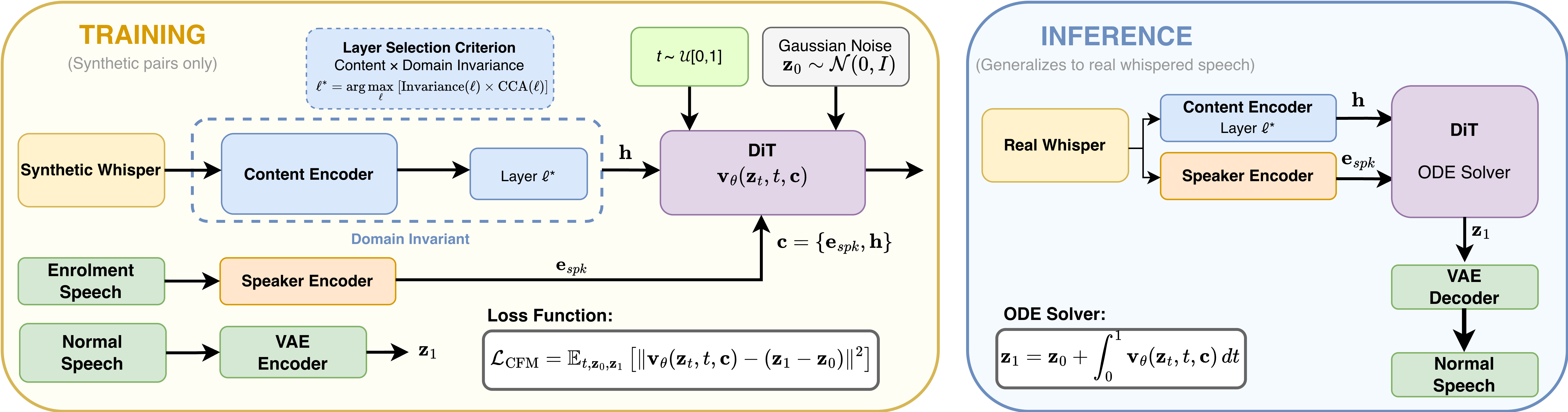}\vspace*{-10pt}
    \caption{FlowW2N pipeline. \textbf{Left (Training):} The DiT learns a velocity field $\mathbf{v}_\theta(\mathbf{z}_t, t, \mathbf{c})$ conditioned on domain-invariant content features $\mathbf{h}$ from a content encoder (layer $\ell^*$) and speaker embedding $\mathbf{e}_{\text{spk}}$, where $\mathbf{c} = \{\mathbf{e}_{\text{spk}}, \mathbf{h}\}$ represents the conditioning set of content and speaker. Training uses only synthetic whisper-normal pairs. \textbf{Right (Inference):} Starting from Gaussian noise, the ODE is integrated to obtain $\mathbf{z}_1$, which is decoded to normal speech. Domain invariance of content features enables generalization to real whispered speech.}\vspace*{-15pt}
    \label{fig:method}
\end{figure*}

We propose FlowW2N, a conditional flow matching approach that sidesteps alignment challenges entirely. Our work is motivated by two observations: (1) synthetic whispered-normal pairs are perfectly aligned by construction, eliminating the phoneme boundary blur problem during training; and (2) if the conditioning features are \textit{domain-invariant}; that is, features are similar whether extracted from synthetic or real whispered, then a model trained exclusively on synthetic data can generalize to real whispered speech at inference. This reframes the W2N from \textit{learning temporal alignment} to \textit{selecting an appropriate conditioning representation}, a substantially simpler task. Figure~\ref{fig:method} illustrates the pipeline: during training, the Diffusion Transformer (DiT) learns a velocity field conditioned on domain-invariant content features and speaker embeddings using only synthetic pairs; at inference, the model generalizes to real whispered speech through the invariance of the conditioning signal.

Our contributions are: (i)~we explore conditional flow matching to W2N, achieving state-of-the-art WER on CHAINS and wTIMIT with only 10 inference steps; (ii)~we train exclusively on synthetic data with domain-invariant conditioning, requiring no real paired recordings; and (iii)~we propose a layer selection criterion balancing content informativeness and cross-domain invariance.

\section{Method}
\label{sec:method}

\subsection{FlowW2N Architecture}
\label{sec:architecture}

We operate in a latent space using a fully-convolutional VAE based on the Oobleck encoder-decoder from \texttt{stable-audio-tools}~\cite{evans2024fast}. It compresses waveforms $\mathbf{s} \in \mathbb{R}^{T}$ into latent representations $\mathbf{z} = \mathcal{E}(\mathbf{s}) \in \mathbb{R}^{D \times L}$, where $D{=}64$ is the latent dimension and $L = T/r$ for compression ratio $r$ at $\approx$15.6\,Hz frame rate. The decoder $\mathcal{D}$ reconstructs waveforms as $\hat{\mathbf{s}} = \mathcal{D}(\mathbf{z})$. The VAE is trained with multi-resolution STFT loss, multi-scale discriminator losses, and KL divergence regularization exclusively on normal speech from HiFi-TTS-2 for 80,000 steps and batch size of 256.

Flow matching~\cite{lipman2022flowmatching, liu2022rectified} learns a velocity field $\mathbf{v}_\theta(\mathbf{z}, t)$ that defines an ordinary differential equation (ODE):
\begin{equation}
    \frac{d\mathbf{z}}{dt} = \mathbf{v}_\theta(\mathbf{z}, t)
\end{equation}
Integrating this ODE from $t{=}0$ to $t{=}1$ transports samples from a source distribution to a target distribution. The conditional flow matching (CFM) objective~\cite{lipman2022flowmatching} trains the network using sample pairs. Given $\mathbf{z}_0$ (source) and $\mathbf{z}_1$ (target), the optimal transport interpolation $\mathbf{z}_t = (1 - t)\mathbf{z}_0 + t\mathbf{z}_1$ with constant target velocity $\mathbf{v}_t = \mathbf{z}_1 - \mathbf{z}_0$ and $t \sim \mathcal{U}[0,1]$ yields the loss:
\begin{equation}
    \mathcal{L}_{\text{CFM}} = \mathbb{E}_{t, \mathbf{z}_0, \mathbf{z}_1} \left[ \lVert \mathbf{v}_\theta(\mathbf{z}_t, t) - (\mathbf{z}_1 - \mathbf{z}_0) \rVert^2 \right]
    \label{eq:cfm_loss}
\end{equation}

\inlinepar{Paired Flow Matching Limitation.}
\label{sec:paired_limitation}
A natural approach for W2N sets $\mathbf{z}_0 = \mathcal{E}(\mathbf{s}_{\text{whisper}})$ and $\mathbf{z}_1 = \mathcal{E}(\mathbf{s}_{\text{normal}})$. However, this assumes temporal alignment. In W2N, speakers exhibit different speaking rates between whispered and normal speech~\cite{Morris2002ReconstructionOS}. When frame $i$ of $\mathbf{z}_0$ and $\mathbf{z}_1$ correspond to different phonemes, the interpolation produces acoustically incoherent samples, corrupting the velocity field supervision.

\inlinepar{Gaussian Prior with External Conditioning.}
To circumvent alignment requirements, we adopt a formulation where the source distribution is Gaussian noise: $\mathbf{z}_0 \sim \mathcal{N}(\mathbf{0}, \mathbf{I})$. Generation is guided by a conditioning signal $\mathbf{c}$ extracted from the whispered input. The velocity field has the form $\mathbf{v}_\theta(\mathbf{z}_t, t, \mathbf{c})$, and the training objective is:
\vspace*{-2pt}\begin{equation}
    \mathcal{L} = \mathbb{E}_{t, \mathbf{z}_0 \sim \mathcal{N}(\mathbf{0}, \mathbf{I}), \mathbf{z}_1} \left[ \lVert \mathbf{v}_\theta(\mathbf{z}_t, t, \mathbf{c}) - (\mathbf{z}_1 - \mathbf{z}_0) \rVert^2 \right]
    \label{eq:cond_loss}\vspace*{-2pt}
\end{equation}
Since $\mathbf{z}_0$ is unstructured noise, the interpolation path no longer blends misaligned phonetic content. The conditioning signal $\mathbf{c}$ provides the necessary information to guide generation toward the correct linguistic content.

\inlinepar{Model Architecture.}
The FlowW2N generative model is a Diffusion Transformer (DiT)~\cite{peebles2023scalable} with 24 transformer blocks, adapted from \cite{evans2024fast}, with timestep $t$ injected via adaptive layer normalization (AdaLN). We explore two conditioning mechanisms: \textbf{cross-attention} (-ca), where each transformer block attends to $\mathbf{c}$, and \textbf{prepending} (-p), where $\mathbf{c}$ is concatenated to the latent sequence. A frozen ECAPA-TDNN\footnote{\href{https://huggingface.co/speechbrain/spkrec-ecapa-voxceleb}{speechbrain/spkrec-ecapa-voxceleb}}~\cite{desplanques2020ecapa} speaker encoder provides speaker embeddings $\mathbf{e}_{\text{spk}}$ via AdaLN, while content features $\mathbf{h}$ from the Whisper encoder are introduced through cross-attention or prepending. The final conditioning signal is $\mathbf{c} = \{\mathbf{e}_{\text{spk}}, \mathbf{h}\}$.

At inference, we sample $\mathbf{z}_0 \sim \mathcal{N}(\mathbf{0}, \mathbf{I})$, and solve the ODE:\vspace*{-2pt}
\begin{equation}
    \mathbf{z}_1 = \mathbf{z}_0 + \int_0^1 \mathbf{v}_\theta(\mathbf{z}_t, t, \mathbf{c}) \, dt\vspace*{-2pt}
\end{equation}
 The final waveform is $\hat{\mathbf{s}} = \mathcal{D}(\mathbf{z}_1)$, $N=10$ steps, Euler integration.

\subsection{Domain Invariance and Layer Selection}
\label{sec:invariance_layer}

For our synthetic-only training strategy to succeed, the conditioning signal must be \textit{domain-invariant}: features from synthetic whisper must closely match those from real whisper to enable generalization. We evaluate domain invariance on the CHAINS dataset, which provides paired recordings of identical utterances in normal and whispered speech. For each utterance, we generate a synthetic whisper, yielding triplets of (normal, real whisper, synthetic whisper) with matched content. We extract word-level representations using forced alignment via WhisperX~\cite{bain2023whisperx} with ground-truth transcripts and compute Pearson correlation between matched representations.

\begin{figure}[t]
    \centering
    \includegraphics[width=0.75\columnwidth]{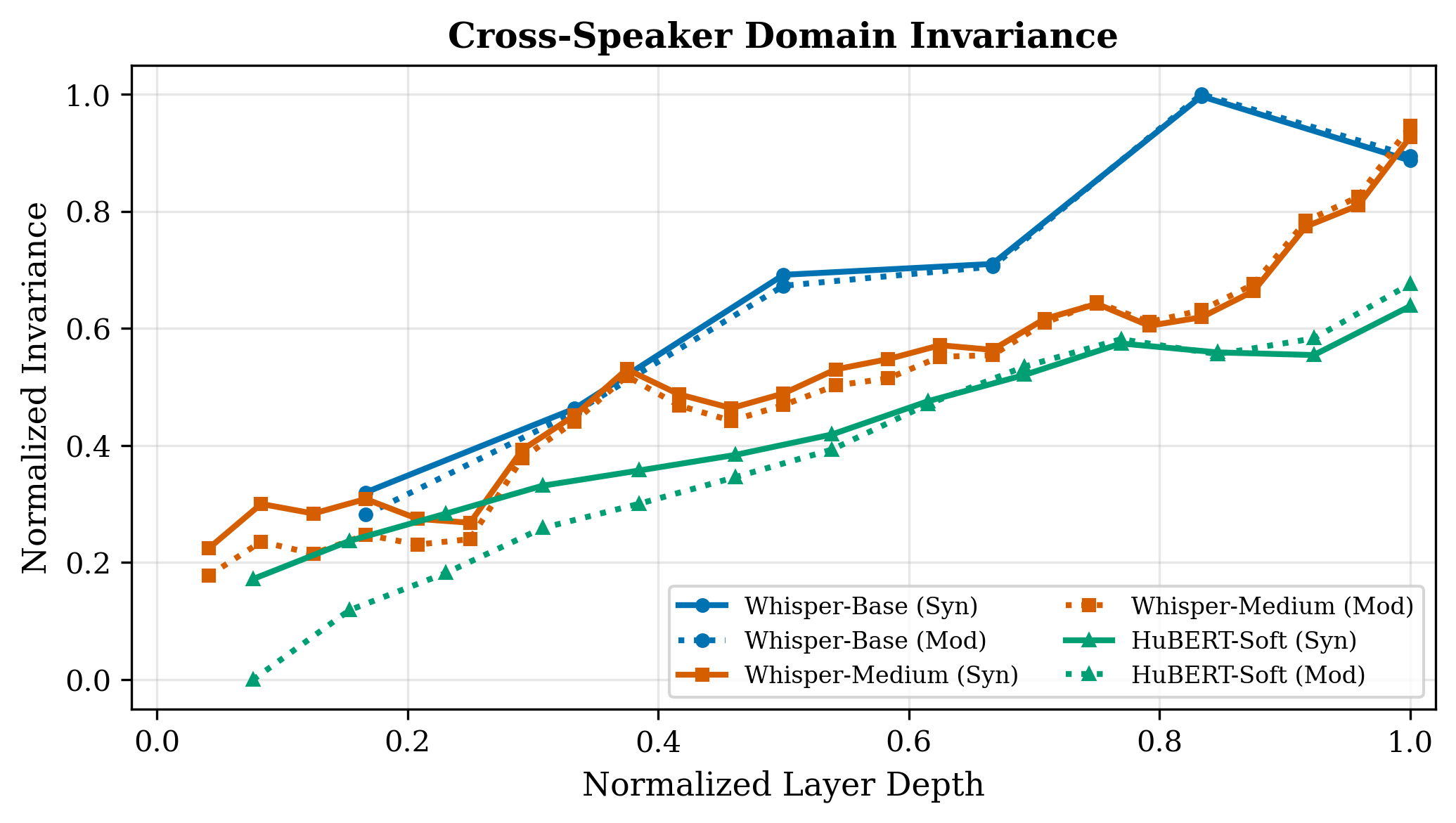}\vspace*{-10pt}
    \caption{Domain invariance analysis. Synthesis Gap (Syn): synthetic vs.\ real whisper; Modality Gap (Mod): real whisper vs.\ normal.}
    \label{fig:invariance}\vspace*{-15pt}
\end{figure}

We measure two gaps: (1)~\textbf{Synthesis Gap}: similarity between synthetic and real whisper features; (2)~\textbf{Modality Gap}: similarity between normal speech and real whisper features. For our strategy to succeed, the synthesis gap must be small. Figure~\ref{fig:invariance} presents layer-wise results. Whisper features achieve consistently higher correlation than HuBERT Soft~\cite{hsu2021hubert,van2022comparison} across all layers, with later layers exceeding 0.90 Pearson correlation. Notably, the synthesis gap and modality gap curves nearly overlap for Whisper embeddings, confirming they are sufficiently domain-invariant for our synthetic-only training strategy.

\begin{figure}[t]
    \centering
    \includegraphics[width=\columnwidth]{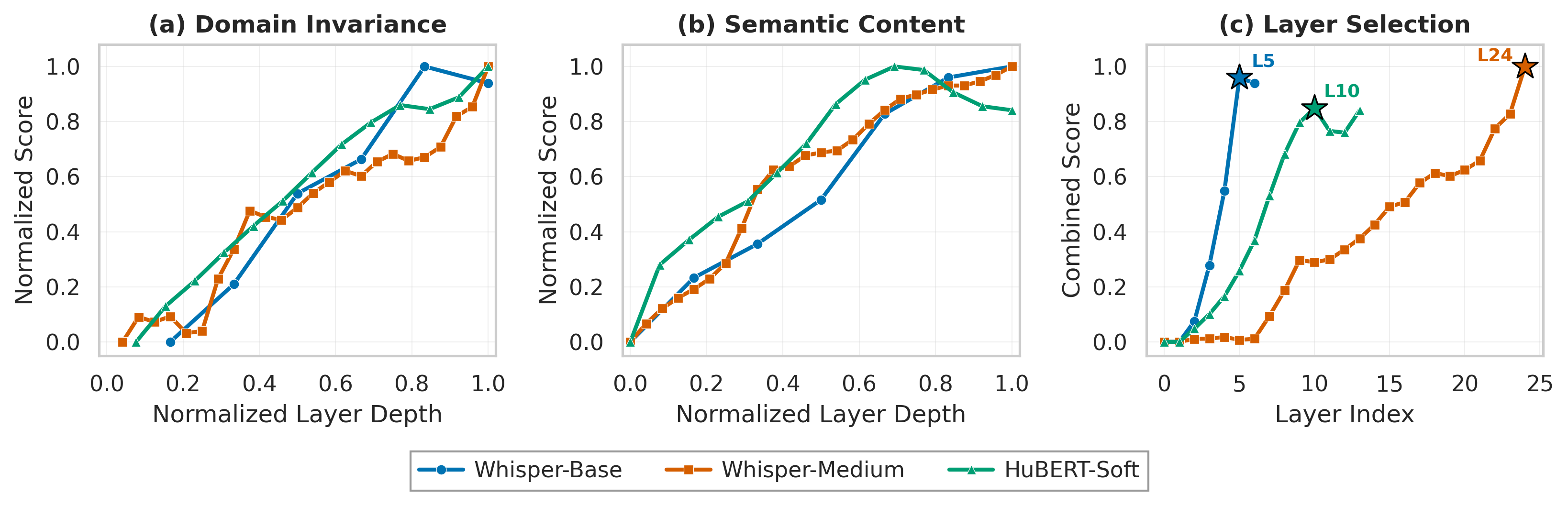}\vspace*{-10pt}
    \caption{Layer selection analysis. Left: Synthesis Gap (invariance). Center: CCA with word identity. Right: Proposed combined score (invariance $\times$ semantic). Stars mark optimal layers selected.}
    \label{fig:layer_selection}\vspace*{-15pt}
\end{figure}

\inlinepar{Layer Selection.}
\label{sec:layer_selection}
Encoder layers exhibit different properties~\cite{pasad-analysis}: early layers capture acoustic details while later layers encode abstract linguistic content. We propose selecting the optimal layer $\ell^*$ by combining: (i)~\textit{Content Informativeness} via CCA between frame-level features and word identity, and (ii)~\textit{Cross-Domain Invariance} via Pearson correlation between synthetic and real whisper features. Both metrics are min-max normalized to $[0,1]$ and combined:
\vspace*{-3pt}\begin{equation}
    \ell^* = \arg\max_\ell \, [ {\text{Invariance}}(\ell) \times {\text{CCA}}(\ell)]
    \label{eq:layer_sel}\vspace*{-5pt}
\end{equation}
Figure~\ref{fig:layer_selection} shows this analysis for HuBERT, HuBERT Soft, Whisper Base, and Whisper Medium. For Whisper Base, layer 5 is optimal ($\ell^*{=}5$), while for HuBERT, layer 10 is optimal ($\ell^*{=}10$). Both components of Eq.~\ref{eq:layer_sel} are grounded in the representation-analysis literature: CCA-based similarity between frame-level features and word identity follows the SVCCA framework for probing what a representation encodes~\cite{raghu2017svcca}, while cross-domain Pearson correlation is the same invariance measure used in Section~\ref{sec:invariance_layer} to establish domain invariance. The criterion is therefore a principled combination of two established measures rather than an ad hoc heuristic.

\section{Experimental Setup}
\label{sec:experimental_setup}

\subsection{Synthetic Whisper Data Generation}
\label{sec:synthetic_data}

To train the FlowW2N model, we generate synthetic whisper-normal pairs from HiFi-TTS-2~\cite{hifitts2} using four techniques sampled with equal probability: (1) LPC-based devoicing\footnote{\href{https://github.com/zeta-chicken/toWhisper}{github.com/zeta-chicken/toWhisper}} that replaces voiced excitation with noise-based excitation; (2) glottal source removal following~\cite{dataaug-whisper-asr} which removes voicing information; (3) formant bandwidth modification~\cite{dataaug-whisper-asr} which increases formant bandwidths to simulate whispered resonances; and (4) the Praat Vocal Toolkit\footnote{\href{https://www.praatvocaltoolkit.com/whisper.html}{praatvocaltoolkit.com/whisper.html}}. Mixing four methods diversifies the synthesis artifacts seen in training; all yield time-aligned pairs by construction.

\subsection{Evaluation Datasets}
\label{sec:eval_datasets}

We evaluate on two real-world whispered speech corpora:

\textbf{wTIMIT}~\cite{lim2010wtimit} contains parallel whispered and normal recordings of TIMIT sentences from 50 speakers. We evaluate on the US dialect test set (1,404 utterances, $\approx$ 2 hours).

\textbf{CHAINS}~\cite{cummins2006chains} contains 36 speakers recorded in multiple speaking styles including whispered and normal speech ($\approx$1,332 whispered utterances, $\approx$ 2.45 hours).

Training data does not contain any real-world whispered data. Therefore, the evaluation data is serving exclusively to measure generalization from synthetic to unseen real whispered speech.

\subsection{Evaluation Metrics}
\label{sec:eval_metrics}

We assess intelligibility using two ASR systems to ensure robustness: (1)~NeMo FastConformer (N), and (2)~Whisper tiny~\cite{radford2023robust} (W). Since Whisper features condition our model's generation, evaluating with Whisper tiny alone could favor our method; we therefore report NeMo FastConformer, an architecturally unrelated ASR, to rule out this bias. We use UTMOS~\cite{saeki2022utmos} and DNSMOS~\cite{dnsmos} to estimate Mean Opinion Scores (MOS) for naturalness and quality, and compute speaker similarity (SpkSim) via Resemblyzer.

To assess phonation reconstruction directly, we extract fundamental frequency (F0) with WORLD (Harvest with StoneMask refinement, 5\,ms hop) and align each converted utterance to the parallel normal recording via Dynamic Time Warping on 25-dimensional MFCCs. We report log-F0 RMSE in semitones over jointly-voiced aligned frames and Pearson correlation of log-F0 contours.

Finally, we conduct a subjective listening study with 25 listeners rating 5 utterances per system on a 5-point MOS naturalness scale (Reference, real Whisper, DistillW2N, QuickVC, and FlowW2N), to address the concern that objective metrics alone cannot establish perceptual quality.

\begin{table*}[t]
\centering
\caption{FlowW2N versus prior methods. WER using NeMo (N) and Whisper tiny (W).
\textcolor{improvement}{Green}: relative improvement over best baseline.}
\label{tab:main_results}
\small
\setlength\tabcolsep{4pt}
\resizebox{0.8\textwidth}{!}{%
\begin{tabular}{@{}ll cc cc cc cc@{}}
\toprule
& & \multicolumn{4}{c}{\textbf{CHAINS}} & \multicolumn{4}{c}{\textbf{wTIMIT}} \\
\cmidrule(lr){3-6} \cmidrule(lr){7-10}
\textbf{Method} & \textbf{Training Data} & \textbf{WER-N\%} $\downarrow$ & \textbf{WER-W\%} $\downarrow$ & \textbf{SpkSim} $\uparrow$ & \textbf{DNSMOS} $\uparrow$ & \textbf{WER-N\%} $\downarrow$ & \textbf{WER-W\%} $\downarrow$ & \textbf{SpkSim} $\uparrow$ & \textbf{DNSMOS} $\uparrow$ \\
\midrule
\multicolumn{10}{l}{\textit{Reference}} \\
\quad Real Whisper & -- & 11.1 & 31.6 & -- & 1.50 & 8.3 & 26.5 & -- & 1.21 \\
\quad Normal Speech & -- & 5.9 & 11.9 & -- & 3.06 &  3.9 & 9.5 & -- & 3.21 \\
\midrule
\multicolumn{10}{l}{\textit{Prior Work}} \\
\quad QuickVC & Real pairs & 24.1 & 36.8 & \textbf{0.668} & \textbf{2.99} & 26.0 & 39.1 & \textbf{0.720} & \textbf{3.04} \\
\quad DistillW2N & Real pairs + Distill & 105 & 201 & 0.576 & 2.20 & 110 & 205 & 0.618 & 2.82 \\
\quad DistillW2N-T & Real pairs + Distill & 37.3 & 54.6 & 0.604 & 2.81 & 37.5 & 61.8 & 0.640 & 3.02 \\
\midrule
\multicolumn{10}{l}{\textit{Ours}} \\
\quad FlowW2N & Synthetic only & \textbf{17.9} \textcolor{improvement}{\scriptsize($\downarrow$26\%)} & \textbf{28.8} \textcolor{improvement}{\scriptsize($\downarrow$22\%)} & 0.631 & 2.93 & \textbf{14.0} \textcolor{improvement}{\scriptsize($\downarrow$46\%)} & \textbf{27.9} \textcolor{improvement}{\scriptsize($\downarrow$29\%)} & 0.700 & 2.91 \\
\bottomrule
\end{tabular}
}\vspace*{-10pt}
\end{table*}

\section{Results}
\label{sec:results}

\subsection{Analysis of Paired Flow Matching}
\label{sec:unconditional_analysis}

We first validate that paired flow matching fails for W2N by evaluating a paired flow matching model (Paired-Base), as described in Section~\ref{sec:architecture}, trained on synthetic HiFi-TTS-2 pairs without external conditioning. Evaluation with CHAINS is done under two conditions: (1)~real whisper input (out-of-domain, OOD), and (2)~synthetic whisper input (in-domain, ID).

As shown in Table~\ref{tab:unconditional_results}, Paired-Base achieves excellent in-domain results: 10.1\% WER, UTMOS 3.46, closely approaching ground truth normal speech quality. However, with real whispered speech (OOD), performance degrades drastically: WER increases to 28.2\% while UTMOS drops to 1.47, nearly identical to the UTMOS value of the real whisper speech (1.42). Subjective listening confirmed that the model effectively acts as an identity function, failing to reconstruct modal phonation, it merely output the input whisper.

To address this, we finetuned Paired-Base on real whisper-normal pairs from wTIMIT, EARS, and whispered class in Expresso~\cite{expresso} dataset. We evaluate two finetuning strategies: direct finetuning on unaligned real pairs (FT-Real), and finetuning with Dynamic Time Warping applied to align the pairs before training (FT-DTW). Both show significant degradation: FT-Real yields 102\% WER, while FT-DTW provides only modest improvement (79.9\% WER), confirming that DTW cannot guarantee the frame-level phonetic coherence required for plausible flow matching trajectories. Table~\ref{tab:unconditional_results} therefore directly tests the claim made in Section~\ref{sec:intro} that paired flow matching is unsuitable for W2N: even with an alignment-aware baseline (FT-DTW), paired training on real data (79.9\% WER) remains far worse than our synthetic-only, externally-conditioned approach (17.9\% WER, Table~\ref{tab:main_results}).

\begin{table}[t]
\centering
\caption{Paired flow matching fails to generalize on real whisper (OOD). Finetuning on misaligned real pairs degrades further.}
\label{tab:unconditional_results}
\small
\setlength\tabcolsep{3pt}
\renewcommand{\arraystretch}{0.85}
\resizebox{0.85\columnwidth}{!}{%
\begin{tabular}{@{}llcccc@{}}
\toprule
\textbf{Model} & \textbf{Input} & \textbf{WER\%} $\downarrow$ & \textbf{UTMOS} $\uparrow$ & \textbf{DNSMOS} $\uparrow$ & \textbf{SpkSim} $\uparrow$ \\
\midrule
\multicolumn{6}{l}{\textit{Reference}} \\
\quad Real Whisper & -- & 11.1 & 1.42 & 1.50 & -- \\
\quad Normal Speech & -- & 5.9 & 3.83 & 3.06 & -- \\
\midrule
\multicolumn{6}{l}{\textit{Uncond. (Synth. Trained)}} \\
\quad Paired-Base & Real (OOD) & 28.2 & 1.47 & 1.43 & 0.522 \\
\quad Paired-Base & \textbf{Synth (ID)} & \textbf{10.1} & \textbf{3.46} & \textbf{3.09} & \textbf{0.832} \\
\midrule
\multicolumn{6}{l}{\textit{Uncond. + FT on Real}} \\
\quad FT-Real & Real (OOD) & 102 & 2.64 & 2.96 & 0.593 \\
\quad FT-DTW & Real (OOD) & 79.9 & 2.92 & 2.97 & 0.613 \\
\bottomrule
\end{tabular}
}\vspace*{-10pt}
\end{table}

\subsection{Comparison with Prior Work}

Table~\ref{tab:main_results} compares the proposed FlowW2N against QuickVC~\cite{quickvc} and DistillW2N~\cite{tan2025distillw2n} on both evaluation corpora. We include DistillW2N, which distills HuBERT Soft representations for efficient W2N conversion, and its teacher-based variant (DistillW2N-T), since the distilled model performs poorly on WER. We re-evaluate DistillW2N on the full CHAINS and wTIMIT test sets; the original paper~\cite{tan2025distillw2n} reports results on a wTIMIT subset, which explains the difference from its published numbers.

FlowW2N achieves the best intelligibility on both datasets and ASR systems. On CHAINS, WER reduces from 24.1\% (QuickVC) to 17.9\% under NeMo. The gains are more pronounced on wTIMIT: 14.0\% vs. 26.0\% (46\% relative). Under Whisper tiny, FlowW2N achieves 27.9\% WER on wTIMIT versus 39.1\% for QuickVC.

Critically, FlowW2N exhibits consistent performance, whereas prior methods show variance. DistillW2N suffers from severe intelligibility degradation (WER $>$ 100\%). While QuickVC achieves marginally higher DNSMOS on CHAINS (2.99 vs. 2.93), we argue that intelligibility measured by WER is the primary objective of W2N conversion. A perceptually pleasant but unintelligible output fails the fundamental task. The DNSMOS gap (0.06 points) is negligible compared to the substantial WER improvements (26--46\% relative).

\subsection{Subjective and Prosody Evaluation}
\label{sec:subj_f0}

Table~\ref{tab:subj_f0} reports a 25-listener MOS naturalness study (5 utterances/system, 5-point scale) and F0 reconstruction on wTIMIT (WORLD F0, DTW-aligned to the normal reference; Section~\ref{sec:eval_metrics}). FlowW2N is rated most natural, reversing the DNSMOS ranking in Table~\ref{tab:main_results} ($p\approx0.06$ vs.\ QuickVC, $p\ll0.001$ vs.\ DistillW2N), and attains the best F0 contour correlation on both wTIMIT and CHAINS (0.27/0.28), while log-F0 RMSE is comparable to QuickVC. Speaker similarity trails QuickVC by 0.02--0.04 (Table~\ref{tab:main_results}); Table~\ref{tab:ablations} shows conditioning choices already recover much of this gap (SpkSim 0.54$\to$0.63).

\begin{table}[t]
\centering
\caption{MOS naturalness (25 listeners) and wTIMIT F0 reconstruction vs.\ the parallel normal recording.}
\label{tab:subj_f0}
\small
\setlength\tabcolsep{2pt}
\renewcommand{\arraystretch}{0.7}
\resizebox{0.7\columnwidth}{!}{%
\begin{tabular}{@{}lccc@{}}
\toprule
\textbf{System} & \textbf{MOS} $\uparrow$ & \textbf{logF0 RMSE} $\downarrow$ & \textbf{F0 corr} $\uparrow$ \\
\midrule
Reference & 4.99 & -- & -- \\
Real Whisper & 3.10 & 11.91 & 0.04 \\
DistillW2N & 1.63 & \textbf{5.38} & 0.12 \\
QuickVC & 3.24 & 5.65 & 0.24 \\
FlowW2N & \textbf{3.37} & 5.88 & \textbf{0.27} \\
\bottomrule
\end{tabular}}\vspace*{-10pt}
\end{table}

\subsection{Ablation Analysis}

Table~\ref{tab:ablations} presents ablation results on CHAINS real whisper input. All models are trained exclusively on synthetic HiFi-TTS-2 data.

\textbf{Conditioning Signal:} C-ASR-p (prepending Whisper Base features for conditioning) achieves 17.0\% WER vs.\ 23.7\% for C-VAE-p (VAE latent embeddings), a 28\% relative improvement, with UTMOS improving from 1.65 to 2.57 (+56\%). ASR features are optimized for recognition across diverse conditions and are largely invariant to the specific acoustic realization, unlike VAE features that retain synthetic whisper-specific details.

\textbf{Speaker Conditioning:} Adding ECAPA-TDNN~\cite{desplanques2020ecapa} speaker embeddings improves output quality across all configurations. UTMOS increases from 2.57 to 3.19 (+24\%), DNSMOS from 2.25 to 2.76 (+23\%), and speaker similarity from 0.54 to 0.60.

\textbf{Content Encoder:}
Among the prepending variants, different content encoders are explored: HuBERT Base, HuBERT Soft layer, and Whisper Base. While HuBERT Soft achieves the highest UTMOS (3.48), we observe differences in WER consistency across ASR systems. All three Nemo encoders yields similar WER (18.60-18.90\%), but under Whisper tiny, HuBERT-based models show higher WER: 36.00\% for HuBERT and 32.30\% for HuBERT Soft, compared to 32.00\% for Whisper Base conditioning. This suggests HuBERT-based models may introduce phonetic distortions that affect intelligibility inconsistently across architectures.  Whisper Base uses only 21M parameters vs.\ HuBERT's 95M (4.5$\times$ more efficient).

\textbf{Cross-Attention vs.\ Prepending:} FlowW2N (C-ASR+Spk-ca) outperforms C-ASR+Spk-p across all metrics: WER improves from 18.7\% to 17.9\% (NeMo) and 32.0\% to 28.8\% (Whisper), with UTMOS from 3.19 to 3.39 and SpkSim from 0.60 to 0.63. Cross-attention allows the model to dynamically attend to relevant conditioning information at each generation step, providing more flexible alignment between content features and generated latents compared to simple concatenation.

\textbf{Layer Selection:} Using Whisper layer 5 ($\ell^*$) instead of layer 12 improves WER from 18.7\% to 16.9\% (10\% relative, NeMo) and 32.0\% to 28.1\% (12\% relative, Whisper tiny), while reducing encoder parameters by 19\%. Similar gains for HuBERT ($\ell^*{=}10$) validate the proposed layer selection criterion. The main-table FlowW2N configuration (Table~\ref{tab:main_results}) is C-ASR+Spk-ca at layer 12; combining cross-attention with the layer-5 selection is a promising direction left to future work.

\begin{table}[t]
\centering
\caption{Ablation study on CHAINS (real whisper input). All models trained on synthetic data only. WER reported for NeMo (N) and Whisper tiny (W).}
\label{tab:ablations}
\small
\setlength\tabcolsep{2.5pt}
\resizebox{\columnwidth}{!}{%
\begin{tabular}{@{}lcccccc@{}}
\toprule
& \multicolumn{2}{c}{\textbf{WER\%} $\downarrow$} & & & \\
\cmidrule(lr){2-3}
\textbf{Configuration} & N & W & \textbf{UTMOS} $\uparrow$ & \textbf{SpkSim} $\uparrow$ & \textbf{DNSMOS} $\uparrow$ \\
\midrule
\multicolumn{6}{l}{\textit{Reference}} \\
\quad Real Whisper & 11.1 & 31.6 & 1.42 & -- & 1.50 \\
\quad Normal Speech & 5.9 & 11.9 & 3.83 & -- & 3.06 \\
\midrule
\multicolumn{6}{l}{\textit{Conditioning Signal (no speaker emb.)}} \\
\quad C-VAE-p & 23.7 & -- & 1.65 & 0.538 & 1.64 \\
\quad C-ASR-p & 17.0 & -- & 2.57 & 0.542 & 2.25 \\
\midrule
\multicolumn{6}{l}{\textit{+ Speaker Embedding (prepending)}} \\
\quad C-HuBERT+Spk-p & 18.6 & 36.0 & 2.87 & 0.581 & 2.65 \\
\quad C-HuBERT-Soft+Spk-p & 18.9 & 32.3 & \textbf{3.48} & 0.601 & \textbf{2.96} \\
\quad C-ASR+Spk-p & 18.7 & 32.0 & 3.19 & 0.601 & 2.76 \\
\midrule
\multicolumn{6}{l}{\textit{+ Speaker Embedding (cross-attention)}} \\
\quad FlowW2N (C-ASR+Spk-ca) & 17.9 & 28.8 & 3.39 & \textbf{0.631} & 2.93 \\
\midrule
\multicolumn{6}{l}{\textit{+ Layer Selection (C-ASR+Spk-p, $\ell^*$=5)}} \\
\quad Layer 5 & \textbf{16.9} \textcolor{improvement}{\scriptsize($\downarrow$10\%)} & \textbf{28.1} \textcolor{improvement}{\scriptsize($\downarrow$12\%)} & 3.41 & 0.596 & 2.89 \\
\multicolumn{6}{l}{\textit{+ Layer Selection (C-HuBERT+Spk-p, $\ell^*$=10)}} \\
\quad HuBERT Layer 10 & 17.5 \textcolor{improvement}{\scriptsize($\downarrow$6\%)} & 35.8 \textcolor{improvement}{\scriptsize($\downarrow$1\%)} & 2.84 & 0.580 & 2.61 \\
\bottomrule
\end{tabular}\vspace*{-10pt}
}
\end{table}
\vspace*{-10pt}
\section{Conclusion}
\label{sec:conclusion}

We presented FlowW2N, a conditional flow matching approach for W2N conversion that addresses temporal misalignment by training on synthetic pairs and conditioning on domain-invariant ASR embeddings. Analysis shows that Whisper encoder features meet this invariance criterion, outperforming VAE and HuBERT representations, and introduced a layer selection criterion balancing content and cross-domain generalization. FlowW2N achieves state-of-the-art intelligibility on CHAINS and wTIMIT, reducing WER by 26--46\% relative to prior work, with only 10 inference steps and no real paired data. Subjective test and F0-contour analyses confirm the naturalness and prosody of the reconstructed speech.

\vfill\pagebreak

\footnotesize
\bibliographystyle{IEEEbib}
\bibliography{mybib}

\end{document}